\documentclass[fleqn,10pt]{wlscirep}
\usepackage[utf8]{inputenc}
\usepackage[T1]{fontenc}
\title{Opinion-Driven Vaccination and Epidemic Dynamics on Heterogeneous Networks}

\author[1,+]{Anika Roy}
\author[1,+]{Ujjwal Shekhar}
\author[2,*]{Subrata Ghosh}
\author[2]{Tomasz Kapitaniak}
\author[1]{Chittaranjan Hens}
\affil[1]{Centre for Computational Natural Sciences and Bioinformatics, International Institute of Information Technology, Hyderabad 500032, India}
\affil[2]{ Division of Dynamics, Lodz University of Technology, Stefanowskiego 1/15, Lodz 90-924, Lodz, Poland}

\affil[*]{jaba.subrata94@gmail.com}

\affil[+]{these authors contributed equally to this work}

\begin{abstract}
Vaccination campaigns play a pivotal role in controlling infectious diseases. Their success, however, depends not only on vaccine efficacy and availability but also significantly on public opinion and the willingness of individuals to vaccinate. This paper investigates a coupled opinion-epidemic model on heterogeneous networks, where individual opinions influence vaccination probability, and opinions themselves evolve through a combination of peer interaction and local risk perception derived from observed infection rates. Embedding the coupled dynamics in scale-free networks, particularly Barabási–Albert structures, allows us to examine the role of network heterogeneity beyond homogeneous-mixing assumptions. Using Monte Carlo simulations and a semi-analytical microscopic Markov-chain approach, we derive and numerically validate analytical expressions for the critical infection threshold and stable vaccinated population where risk perception dominated peer influence.
Our results show that stronger local risk perception enhances pro-vaccination opinions and suppresses infection, while dominant peer influence can increase long-term infection levels. These findings underscore the importance of accounting for social behavior and network structure when designing effective vaccination and epidemic control strategies.
\end{abstract}
\begin{document}

\flushbottom
\maketitle
%
%
\thispagestyle{empty}


\section*{Introduction}
Despite the strong effectiveness of vaccination strategy for mitigating infectious diseases \cite{alvarez2022spatial}, like  measles, meningitis, and COVID-19, vaccine hesitancy is an escalating public health concern in the context of highly contagious infectious diseases~\cite{wilder2020resurgence}. In particular, responses to vaccination are  strongly shaped by human behavior, including fear of infection, information awareness, adverse opinion, and attitudes toward interventions \cite{perra2011towards,vardavas2007can,dube2013vaccine,de2025interplay,dube2021vaccine,math11143109,wang2016statistical,zhang2010hub,nie2025vaccination,alahmadi2025modelling,small2005clustering,chen2025protect}. As a result, modern public health dynamics increasingly emerge from the coupled interplay between biological contagion processes and social-behavioral feedbacks \cite{funk2010modelling,wang2016statistical,pastor2001epidemic,pastorsatorras2015epidemic,zhai2025impact,roy2024impact}.
Thus, understanding these intricate feedback loops between opinion formation and epidemic spread is paramount for developing effective public health strategies, especially in an era susceptible to misinformation and polarized social discourse \cite{ruan2020malicious}. 
\par
Opinion formation is a multifaceted process influenced by social interaction, personal beliefs, and environmental cues \cite{castellano2009statistical,starnini2025opinion,urena2019review,lorenz2007continuous,wu2012impact,hu2025expressed,ren2024two}. The collective opinion of a large population can lead to emergent behavior in diverse domains such as opinion polarization and depolarization in elections \cite{ojer2023modeling,berelson1954voting}, financial market crashes \cite{granha2022opinion}, the formation of social groups \cite{altafini2012dynamics}, and the spread of rumor in community \cite{baumann2020modeling,nekovee2007theory,saha2025pattern}. 
Researchers have long studied opinion dynamics \cite{oestereich2019three, biswas2011phase,suchecki2025biswas,starnini2025opinion} using two broad classes of models: discrete models, where opinions take distinct values, and bounded confidence models, where opinions are continuous within a bounded domain. Well-known discrete models include the voter model \cite{clifford1973model,holley1975ergodic,choi2025analysis}, the majority rule model \cite{galam2002minority,nguyen2020dynamics}, and the Sznajd model \cite{sznajd2000opinion,karan2017modeling,shang2021opinion}. Prominent continuous models include the Deffuant model \cite{deffuant2000mixing} and the Hegselmann–Krause model \cite{rainer2002opinion}. Discrete models are suited to binary choices, whereas continuous bounded models better capture gradual opinion evolution.

In the context of epidemics,  models suggest how  human decision-making regarding vaccination uptake is shaped by  opinions and beliefs spread through social networks \cite{baumann2020modeling,ni2023heterogeneous,xu2024discrete,alvarez2017epidemic,chen2023dynamics,qian2025information,bhowmick2022analysis}. For instance, in \cite{alvarez2017epidemic,jankowski2022role}, two layers are considered in which the updates of opinion layers are controlled by the probabilistic interaction of two interacting nodes. The second layer represents the true physical contact that drives the spreading. In a multilayer setup, it can be shown that proper diffusion with human awareness, adverse opinions, and associated intervention strategies  based on opinion  shape   the epidemics spreading\cite{chen2023dynamics,de2025interplay,fang2023coevolution,math11143109,da2019epidemic,wang2020epidemic}. 
The  work by Pires et al.\  \cite{pires2018sudden,pires2017dynamics} introduced a coupled opinion-epidemic model that elegantly demonstrated how opinion-driven vaccination can lead to abrupt, first-order phase transitions in epidemic prevalence. This work highlighted the potential for nonlinear responses in public health outcomes based on social factors, and opinion dynamics has been used as discrete dynamics. Similar types of  studies have been reported recently that include  agent-level heterogeneity \cite{reitenbach2024coupled,teslya2022effect}, and adaptive network rewiring \cite{oestereich2020hysteresis,oestereich2023optimal}.
Most recently, in ~\cite{de2025interplay}, the  dynamic vaccination campaign is modelled  as a binary state—inactive (hesitant) H or active (pro-vaccine) A—with transitions governed by a Watts-Granovetter-style \cite{watts2007influentials} threshold rule within an age-structured multilayer network coupled to an SIR epidemic. All of these studies, however, ignore the combined impact of neighbors' opinions and infections on the node's vaccination uptake rate, and the opinion of each node does not have a dynamical evaluation (except \cite{pires2018sudden,pires2017dynamics}). However, in \cite{pires2018sudden,pires2017dynamics} the opinion dynamics are considered as purely homogeneous, and the connectivity is either fully connected,
or the degree distribution follows a bounded shape.

Despite this reliance on homogeneous mixing, where interactions are treated as random across the entire population, real-world social networks are markedly heterogeneous, exhibiting scale-free degree distributions, community structure, and clustering \cite{barabasi1999emergence,albert2002statistical,watts1998collective,barrat2008dynamical,hens2019spatiotemporal,ghosh2025universal,kundu2017transition,ghosh2021optimal,ji2023signal}. These structural features play a crucial role in shaping both epidemic dynamics and the spread of vaccination attitudes. Yet most coupled opinion-epidemic models overlook these complexities, limiting their ability to capture realistic contagion and behavioural processes.
{ Note that vaccination games capture how individuals compare the costs and benefits of vaccinating versus remaining susceptible, providing a simple way to embed behavior into epidemic models \cite{chang2020game,iwamura2018realistic,satapathi2025game,wang2024effect,wang2025impact}.  Opinion-dynamics models offer a complementary lens, showing how interactions across social networks can trigger abrupt shifts, stable divisions, or polarized communities around vaccine uptake \cite{castellano2009statistical,pires2018sudden,de2025interplay,pires2017dynamics,10.1098/rspb.2010.1107}. 
	
	\begin{figure}[ht]
		\centering
		\includegraphics[width=0.6\linewidth]{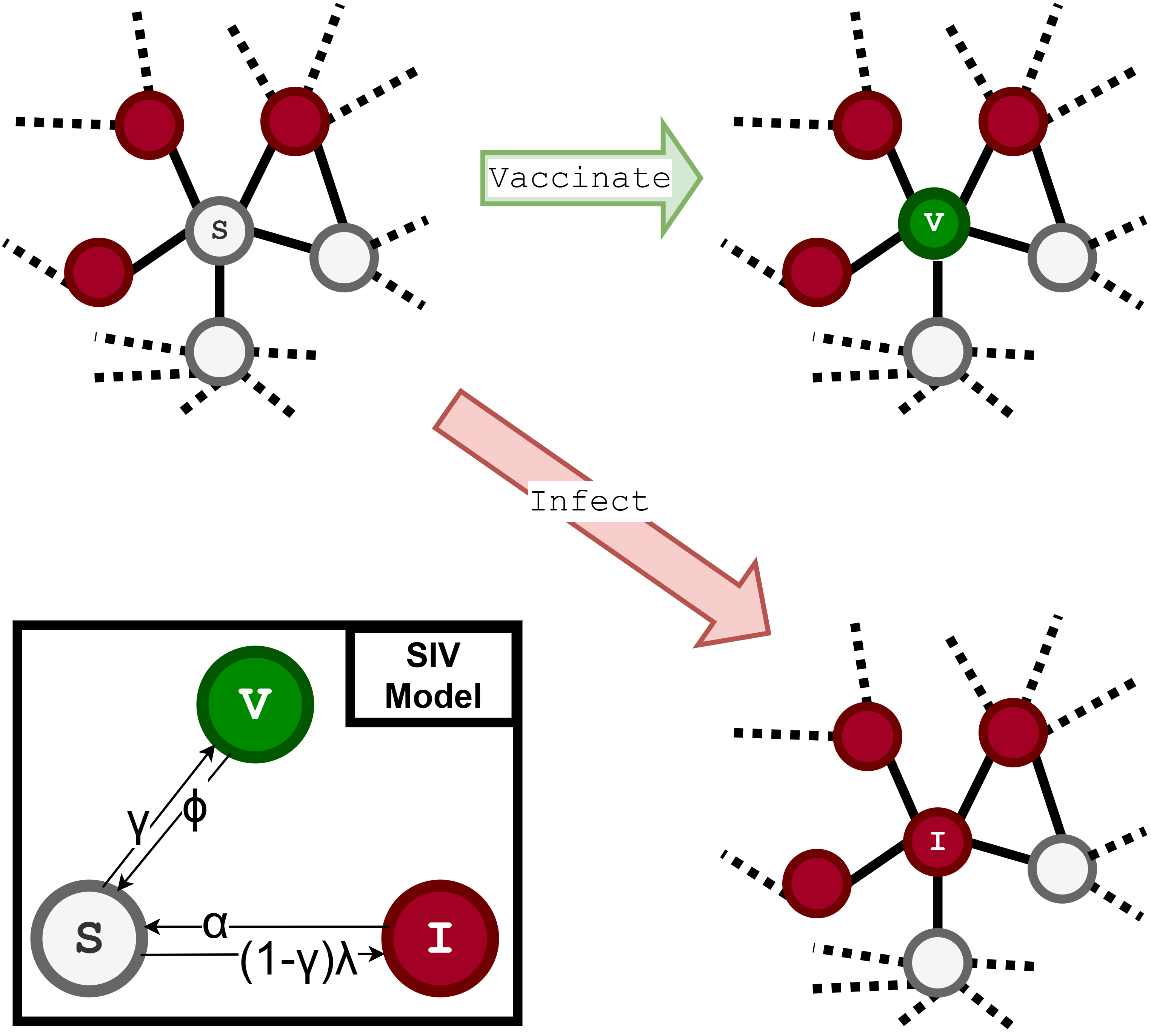}
		\caption{Schematic representation of the coupled SIV model with opinion-dependent vaccination. Susceptible ($S$) individuals can become infected ($I$) or vaccinated ($V$). Infected nodes recover with rate $\alpha$, and vaccinated nodes lose immunity with rate $\phi$. Opinion $O_i$ influences vaccination probability $\gamma_i$.} 
		\label{fig:siv-schematic}
	\end{figure}
	
	\par In this work, we investigate an opinion-driven susceptible-infected-vaccinated (SIV) model in which infected individuals return to susceptibility through a latent recovery stage, and vaccinated individuals gradually lose immunity (see the Fig.\ \ref{fig:siv-schematic}). The framework incorporates two key features:
	(i) the vaccination rate evolves dynamically in response to changes in public opinion, and
	(ii) individual opinions are shaped by peer influence and by risk perception derived from the local network environment. Consequently, the network structure, neighborhood opinions, and infection states collectively govern the opinion update process at each node. Throughout this paper, the terms `nodes' and `individuals' are used interchangeably.
	
	To move beyond homogeneous mixing assumptions, we embed these coupled dynamics within heterogeneous contact networks, with particular emphasis on Barab\'{a}si-Albert (BA) scale-free structures. This approach enables us to analyze how network heterogeneity combined with peer pressure and localized risk perception affects epidemic thresholds and shapes the steady-state behavior of both the opinion dynamics and the epidemic spreading process \cite{GLAUBITZ2023e19094}. We formulate a node-level opinion update mechanism that depends on the previous opinion of individuals, the aggregate opinion of its neighbors, and the local infection level. By interpreting  the scaled opinion value of each node as its vaccination probability, we seamlessly couple the opinion process with the susceptible-infected-vaccinated (SIV) dynamics, yielding a unified framework that links behavioral responses with disease spread. This integration reveals how structural features of the network profoundly shape both epidemic propagation and opinion formation.
	
	A persistent challenge in such coupled dynamics is the derivation of analytical expressions that support numerical findings. Several approaches have been employed, including heterogeneous mean-field approximations~\cite{pastor2001epidemic,pastorsatorras2015epidemic,wang2017unification}. A decade ago, microscopic Markov-chain approach (MMCA) frameworks were developed where discrete-time evolution equations can be written for the entire system, and one can derive analytical insights such as epidemic thresholds directly from the model’s stochastic rules~\cite{granell2013dynamical,gomez2015abrupt,arenas2020modeling,gomez2010discrete}. These methods offer promising pathways to quantify the interplay between network structure, opinion dynamics, and epidemic processes.
	Our primary objective is to investigate the interplay between opinion dynamics and epidemic progression, compare the key factors that drive disease spread, and identify intervention strategies capable of mitigating or suppressing the epidemic. To this end, we employ a dual methodology: MC simulations to capture stochastic behavior at the microscopic level, and a semi-analytical discrete model to provide macroscopic insight. We derive analytical expressions for the critical infection thresholds and validate them numerically, revealing phenomena such as phase transitions, bistability, and crucially a strong consistency between the stochastic simulations and the semi-analytical predictions. This agreement highlights the reliability and practical value of the discrete modeling framework in analyzing complex socio-epidemic systems.}
Our results indicate that increasing risk perception driven by local infection status promotes favorable attitudes toward vaccination, leading to higher vaccine uptake and better infection control. In contrast, stronger peer influence tends to amplify infection levels over longer time scales. Interestingly, the characteristics of the transitions change based on the system parameters: infection dynamics can exhibit smooth or mixed transitions, whereas vaccine uptake and opinion formation show discontinuous like transitions. This study also includes an analytical solution for the critical infection threshold. In addition, for the regime where risk perception dominates peer influence, we derive the stable vaccinated population and verify the results through numerical simulations. Additionally, we validated our results against two real-world online social networks to demonstrate the robustness and practical applicability of the proposed framework in real-world settings.

\section{Model Description}
We use  the coupled opinion-epidemic framework, building upon previous work \cite{pires2018sudden}, to incorporate structured populations represented by networks. Each individual (node) in the network exists in one of three epidemic states: Susceptible ($S$), Infected ($I$), or Vaccinated ($V$). In addition to their epidemic state, each node also possesses a continuous opinion variable $O_i(t)$, which dynamically modulates their individual probability of vaccination. The feedback mechanisms between epidemic dynamics (infection and recovery) and opinion dynamics (peer influence and risk perception) are central to determining the long-term outcomes of the system. The SIV model and an overview of the dynamics of the model has been shown in Fig.\  \ref{fig:siv-schematic}. \par
{\it Epidemic Dynamics.}
The epidemic subsystem is modeled as a discrete-time stochastic process occurring on a network represented by an adjacency matrix $A$. For a population of $N$ individuals (nodes), the state of node $i$ evolves according to the following rules per time step:
(i) A susceptible individual ($S$) can transition to a \emph{vaccinated} state ($V$) with a probability $\gamma_i(t)$. This probability is directly influenced by their current opinion $O_i(t)$.
(ii) A susceptible individual ($S$) can also transition to an \emph{infected} state ($I$) with probability $(1-\gamma_i(t))\lambda$. This probability is determined by their exposure to infected neighbors. The total infection probability for a susceptible individual with $p$ infected neighbors is $(1-\gamma_i(t))(1-(1-\lambda)^p)$~\cite{zhang2010hub}.
(iii) An infected individual ($I$) recovers and returns to the susceptible state ($S$) with a constant probability $\alpha$.
(iv)    A vaccinated individual ($V$) loses immunity and returns to the susceptible state ($S$) with a constant probability $\phi$ per discrete time step, regardless of the time elapsed since vaccination.
\par 
{\it Opinion Dynamics.}
Each individual $i$ maintains a continuous opinion $O_i(t)$ that lies within the bounded interval $[-1,1]$. Negative opinion values (e.g., $O_i=-1$) represent anti-vaccination sentiment, while positive values (e.g., $O_i=1$) indicate pro-vaccination sentiment. Opinions evolve in discrete time steps driven by two primary mechanisms:
\par \textit{Peer Influence ($\epsilon$):} Individuals tend to converge their opinions with those of their neighbors. The parameter $\epsilon$ controls the strength of this social conformity, driving local consensus or shifts in sentiment.
\par \textit{Risk Perception ($\omega$):} Opinions also shift based on the perceived risk of infection from locally observed infected neighbors. The parameter $\omega$ quantifies an individual's sensitivity to this risk, pushing opinions towards pro-vaccination when infection is prevalent in their vicinity.
The update rule for the opinion $O_i(t)$ is given by:
\begin{equation}
	O_i(t+1) = O_i(t) 
	+ \frac{\epsilon}{K_i}\sum_{j=1}^N A_{ij} O_j(t) 
	+ \frac{\omega}{K_i}\sum_{j=1}^N A_{ij} I_j(t),
	\label{eq:opinion-update}
\end{equation}
where $K_i = \sum_{j=1}^N A_{ij}$ is the degree of node $i$. We clip the value of the result to ensure that opinions remain within the $[-1,1]$ range, consistent with bounded-confidence models \cite{deffuant2000mixing,castellano2009statistical}.
Opinion directly modulates the probability of vaccination for susceptible individuals:
\begin{equation}
	\gamma_i(t) = \frac{1 + O_i(t)}{2}, \qquad 0 \leq \gamma_i(t) \leq 1.
	\label{eq:vaccination-prob}
\end{equation}
This formulation ensures that individuals with a strongly pro-vaccine opinion ($O_i=1$) will vaccinate with a probability of 1, while those with a strongly anti-vaccine opinion ($O_i=-1$) will not vaccinate (probability 0). Intermediate opinions yield proportional vaccination propensities, reflecting observed behavioral heterogeneity in real-world vaccination uptake \cite{bauch2013social}.
\par 
{\it Network Topology. }
We focus on Barab\'{a}si–Albert (BA) networks \cite{barabasi1999emergence}, which exhibit a scale-free degree distribution $P(k) \sim k^{-\gamma_\text{network}}$. The adjacency matrix $A$ is the mathematical encoding of the generated network's topology, where each entry $A_{ij} = 1$ represents a physical connection between nodes $i$ and $j$ as defined by the edge set, and $A_{ij} = 0$ otherwise. Their heterogeneous structure, comprising a few highly connected hubs and many low-degree nodes, closely mirrors real-world social networks and strongly influences both epidemic spreading and opinion formation \cite{pastorsatorras2015epidemic}. In our simulations, we consider BA networks with $N = 2000$ nodes, $\gamma_\text{network}=3$ and an average degree $\langle k \rangle \approx 12$.

\begin{figure}
	\centering
	\includegraphics[width=0.6\linewidth]{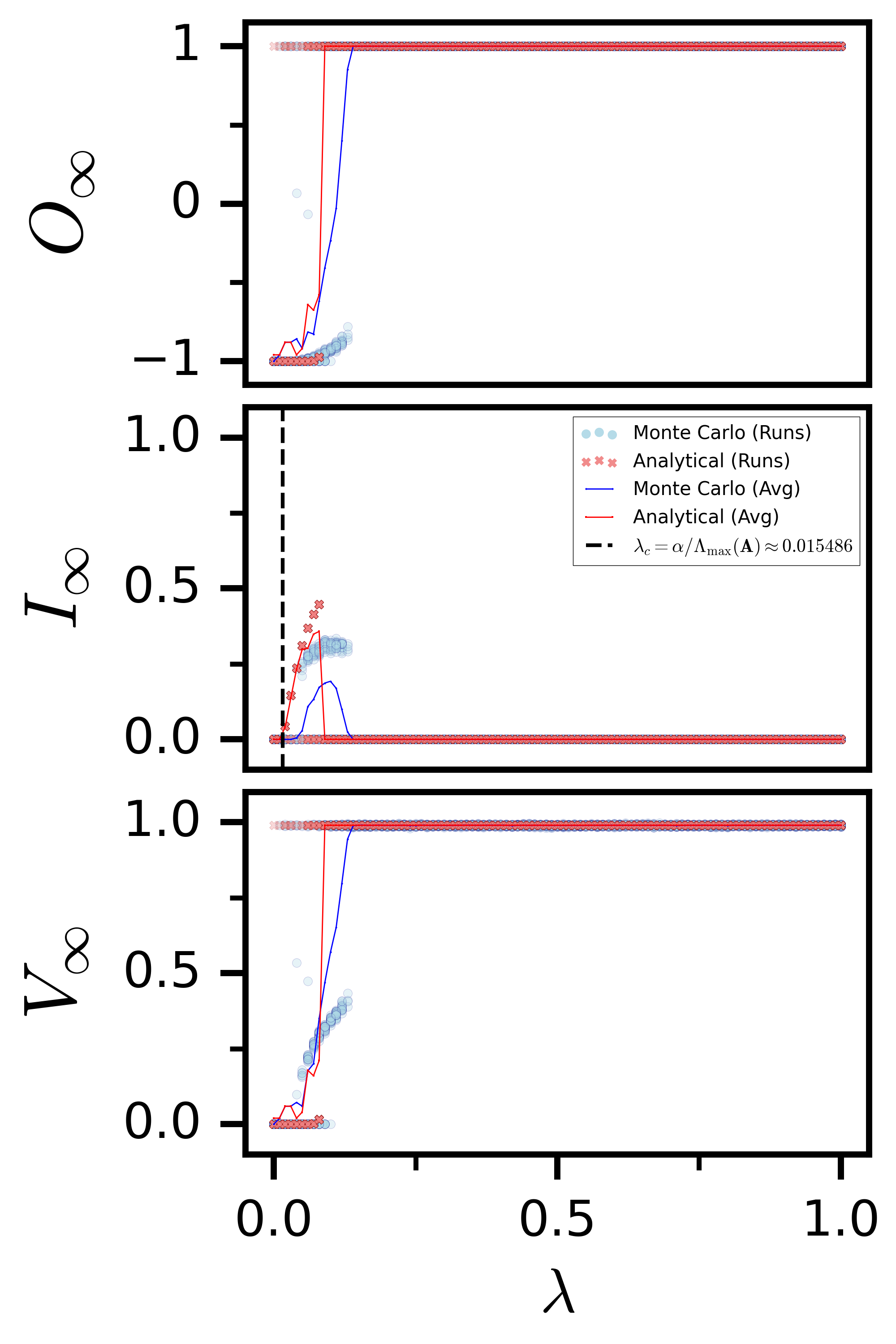} 
	\caption{Steady-state average opinion ($O_{\infty}$), fraction infected ($I_{\infty}$), and fraction vaccinated ($V_{\infty}$) as functions of the infection rate $\lambda$ (with parameters $\alpha = 0.4$, $\epsilon = 0.5$, $\omega = 0.8$, and $\phi = 0.01$). For each $\lambda$, 50 realizations were performed for both the MC and analytical models. Individual MC realizations are shown as light blue dots, with their mean trend indicated by the solid blue line. Similarly, analytical results are represented by light red crosses, and their mean trend by the solid red line. The theoretical infection threshold, $\lambda_c \approx 0.0155$ (derived in Section~\ref{Semi-analytical_MMCA}), is marked on the $I_{\infty}$ plot. For further discussion on the temporal behaviour of these variables, please refer to the appendix \ref{TS_dynamics} (Fig.\ \ref{fig:timeseries_infection}) .
	}
	\label{fig:steady_state_lines}
\end{figure}
\section{Methodology}
\label{methodology}
Our study employs both MC simulations for a stochastic approach and a semi-analytical discrete model for a macroscopic, deterministic approximation.

\subsection{Monte Carlo  simulation in BA networks}
At each discrete time step, every node $i$ updates its opinion and epidemic state. 
The opinion $O_i(t+1)$ is first updated using Eq.~\eqref{eq:opinion-update}, which accounts for both the opinions and infection states of neighboring nodes. 
Next, epidemic transitions occur probabilistically: a susceptible node $i$ vaccinates with probability $\gamma_i(t)$ (Eq.~\eqref{eq:vaccination-prob}) or 
that individual becomes infected from at least one of their neighbors at time $t$ as:
\begin{equation}
	q_i^{\rm SI}(t) = 1 - \prod_{j=1}^N \left( 1 - \lambda A_{ij} I_j(t) \right),
	\label{eq:infection-prob-from-neighbors}
\end{equation}
where $\lambda$ 
is the infection transmission rate, 
$A_{ij}$ 
is an element of the adjacency matrix (1 if individuals $i$ and $j$ are connected, 0 otherwise), and 
$I_j(t)\in [0,1]$ 
denotes the epidemic state of neighbor $j$. This multiplicative form assumes independence of transmission events across neighbors, a standard approach in network epidemic models \cite{pastorsatorras2015epidemic,barrat2008dynamical}. This is the usual contact (parallel) infection term used in discrete-time MMCA for contact-based spreading (see  Subsec.~\ref{Semi-analytical_MMCA}) and matches the per-step infection mechanism used in MC simulations where each infected neighbour independently tries to infect 
$i$. 
Infected nodes recover and return to the susceptible state with probability $\alpha$, while vaccinated nodes lose immunity and become susceptible again with probability $\phi$.

\begin{figure}
	\centering
	\includegraphics[width=0.9\linewidth]{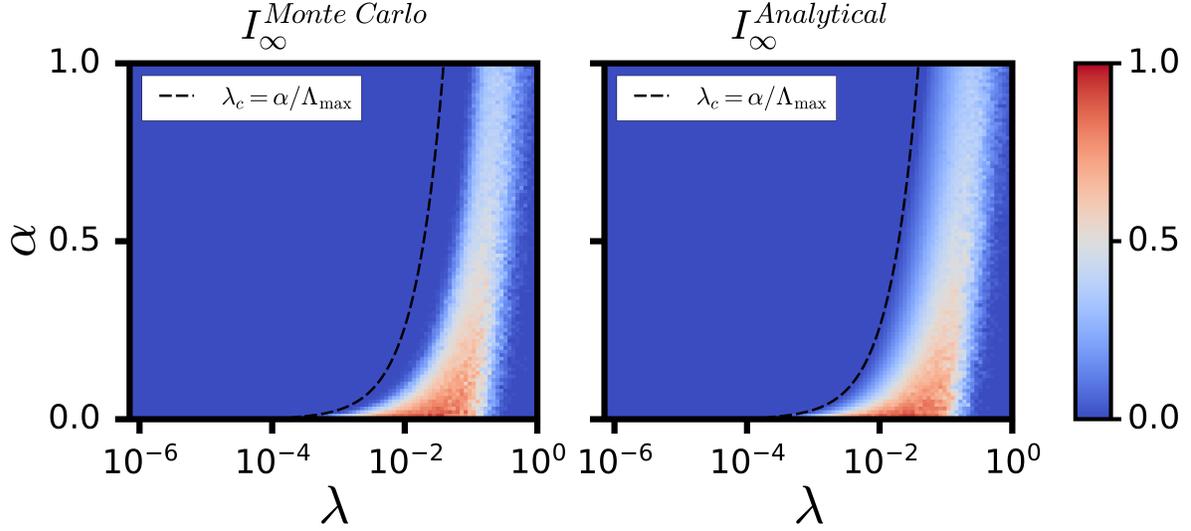}
	\caption{Phase diagrams showing the steady-state infection density ($I_{\infty}$) across varying infection rates ($\lambda$) and recovery rates ($\alpha$). The left panel shows the MC-derived final state, while the right panel uses the semi-analytical model. Both plots fix $\omega=0.8, \phi=0.01, \epsilon=1.0$ 
		The theoretically obtained critical threshold $\lambda_c$ is shown with dashed line. 
	}
	\label{fig:lambda_alpha_phase_plots}
\end{figure}
The simulation tracks the fraction of individuals in each epidemic state ($I$, $V$) and the average opinion ($O = \sum O_i / N$) over time. Note that only tracking ($I$, $V$) is enough to track all of ($S$, $I$, $V$) due to normalization ($S + I + V = N$). We run multiple independent MC simulations (typically 50-100 runs per parameter set) and average the results across these runs to mitigate stochastic fluctuations. 
 For each run, the system executes randomized initialization of opinions using the specified bias fraction ($D$) and infects a random number of individuals ($I_0$). Each timestep is sequentially executed and the opinions and states are updated as shown in Eq.\ \eqref{eq:infection-prob-from-neighbors}. 

\subsection{Semi-Analytical Discrete Model: Microscopic Markov-chain approach (MMCA) framework}
\label{Semi-analytical_MMCA}
The semi-analytical model provides a deterministic description of the average fractions of individuals in each state and their average opinions \cite{pires2018sudden,granell2013dynamical,gomez2015abrupt,arenas2020modeling,gomez2010discrete}.

{
	To derive the discrete-time evolution equation for the infection probability of node~$i$, we condition on its state at time~$t$. The marginal probability that node~$i$ is infected at time $t+1$ is
	\begin{equation}
		I_i(t+1) \;=\; \Pr\!\big[X_i(t+1)=I\big]
	\end{equation}
	
	Using the law of total probability over the possible states of node~$i$ at time~$t$, this becomes
	\begin{equation}
		\begin{split}
			I_i(t+1) = & (\Pr[X_i(t)=I] \cdot \Pr[X_i(t+1)=I \mid X_i(t)=I]) \\
			& + (\Pr[X_i(t)=S] 
			 \cdot \Pr[X_i(t+1) \neq V \mid X_i(t)=S] 
			  \cdot \Pr[X_i(t+1) = I \mid X_i(t)=S \text{ and } X_i(t+1) \neq V])
		\end{split}
		\label{eq:mmca-conditioning}
	\end{equation}

	As described before, an infected node
	remains infected with probability $(1-\alpha)$. A susceptible node avoids vaccination with
	probability $(1-\gamma_i(t))$, where $\gamma_i(t)$ is the opinion-dependent
	vaccination probability. And the probability that a susceptible node~$i$ becomes
	infected by its neighbours during the time step is denoted by
	$q_i^{\mathrm{SI}}(t)$.
	Substituting these quantities into
	Eq.\ \eqref{eq:mmca-conditioning} we capture the  discrete-time equation as
	\begin{equation}
		I_i(t+1)
		=
		I_i(t)\,(1-\alpha)
		\;+\;
		S_i(t)\,\bigl[1-\gamma_i(t)\bigr]\,
		q_i^{\mathrm{SI}}(t),
		\label{eq:I-update}
	\end{equation}
	where the susceptibility probability is
	\begin{equation}
		S_i(t) \;=\; 1 - I_i(t) - V_i(t).
	\end{equation}
	
	\vspace{1em}
	\noindent
	A similar argument gives the marginal update for the vaccination probability.
	A vaccinated node remains vaccinated with probability $(1-\phi)$, where
	$\phi$ is the immunity-waning probability. A susceptible node becomes
	vaccinated with probability $\gamma_i(t)$, giving
	\begin{equation}
		V_i(t+1)
		=
		V_i(t)\,(1-\phi)
		\;+\;
		S_i(t)\,\gamma_i(t).
		\label{eq:V-update}
	\end{equation}

	This expression aligns with the parallel-update mechanism used in MC
	simulations and yields a deterministic, semi-analytical description of the
	average epidemic dynamics on contact networks.
}

The term $q_i^{\rm SI}(t)$ represents the effective probability that a susceptible node $i$ becomes infected, taking into account both its susceptibility based on opinion (i.e., not vaccinating) and the risk of infection from its neighbors. This is crucial for consistency with the MC model. 

The opinion $O_i(t)$ evolves deterministically according to Eq.~\eqref{eq:opinion-update}. For consistency, $I_j(t)$ in the opinion update (Eq.~\eqref{eq:opinion-update}) and in the infection probability (Eq.~\eqref{eq:infection-prob-from-neighbors}) are treated as the probability of node $i$ being infected at time $t$, rather than a binary state. All values ($I_i, V_i, \gamma_i, q_i^{SI}$) are clipped to the range $[0, 1]$ to maintain physical validity. 

{\it Epidemic threshold.}
Next we derive a closed-form critical threshold by analyzing the system under the limit where no individuals are vaccinated, and the infection has reached a steady state. 
We begin by assuming a steady state where the fraction of vaccinated individuals is zero, so $V_i^* = 0$ for each node $i$. For this state to be self-consistent, the probability of vaccination must also be zero, $\gamma_i^* = 0$. This, in turn, implies that all individuals hold a maximally anti-vaccine opinion, $O_i^* = -1$.
Under these conditions ($V_i^*=0, \gamma_i^*=0$), the full steady-state equation for infection simplifies as shown in Eq.\ \eqref{eq:ghosh-start} (infection probability $q_i^{SI}(t)$ in the steady-state limit is denoted as $q_i^*$).
\begin{equation}
	I_i^* = \frac{q_i^*}{\alpha + q_i^*}.
	\label{eq:ghosh-start}
\end{equation}

To find the critical point where an epidemic is about to begin, we analyze the behavior of Eq.\ \eqref{eq:ghosh-start} when the infection level $I_i^*$ is infinitesimally small but non-zero ($I_i^* \to 0$).

First, we rearrange Eq.\ \eqref{eq:ghosh-start} to solve for the recovery term:
\begin{align}
	& I_i^\star(\alpha + q_i^\star) = q_i^\star \nonumber\\
	&\implies \alpha I_i^\star = q_i^\star(1 - I_i^\star).
	\label{eq:rearranged-eq}
\end{align}

Next, we linearize the infection probability term, $q_i^\star$, which is defined as $q_i^\star = 1 - \prod_{j=1}^N (1 - \lambda A_{ij} I_j^\star)$. For very small values of $I_j^*$, this can be approximated as:
\begin{equation*}
	q_i^\star \approx \lambda \sum_{j=1}^N A_{ij} I_j^\star.
\end{equation*}
We now substitute this linear approximation back into Eq.\ \eqref{eq:rearranged-eq}. Since we are at the epidemic threshold, which $I_i^*$ is infinitesimally small, we can take $(1 - I_i^\star) \approx 1$. This gives us:
\begin{equation}
	\alpha I_i^\star \approx \left(\lambda \sum_{j=1}^N A_{ij} I_j^\star\right) (1) = \lambda \sum_{j=1}^N A_{ij} I_j^\star.
	\label{eq:linearized-system}
\end{equation}
This expression represents a system of linear equations for all nodes $i=1, \dots, N$. We can write this system in matrix form, where $\mathbf{I}^\star$ is the column vector of infection probabilities and $\mathbf{A}$ is the adjacency matrix:
\begin{equation*}
	\alpha \mathbf{I}^\star = \lambda (\mathbf{A} \mathbf{I}^\star).
\end{equation*}
This can be rewritten into the standard form of an eigenvalue problem.
For a non-trivial solution (i.e., $\mathbf{I}^\star \neq \mathbf{0}$) to exist, the scalar term $\alpha/\lambda$ must be an eigenvalue of the adjacency matrix $\mathbf{A}$. An epidemics can emerge when the infection rate $\lambda$ is large enough to sustain the spread. This first occurs when the condition is met for the largest eigenvalue (spectral radius) of $\mathbf{A}$, denoted $\Lambda_{\text{max}}(\mathbf{A})$.
Therefore, at the critical infection rate $\lambda = \lambda_c$, we have:
\begin{equation*}
	\frac{\alpha}{\lambda_c} = \Lambda_{\text{max}}(\mathbf{A}).
\end{equation*}
Solving for $\lambda_c$ yields the critical threshold for an epidemic in the absence of vaccination:
\begin{equation}
	\lambda_c = \frac{\alpha}{\Lambda_{\text{max}}(\mathbf{A})}.
	\label{eq:critical-lambda-sis}
\end{equation}

\begin{figure*}[ht]
	\centering
	\includegraphics[width=0.85\textwidth]{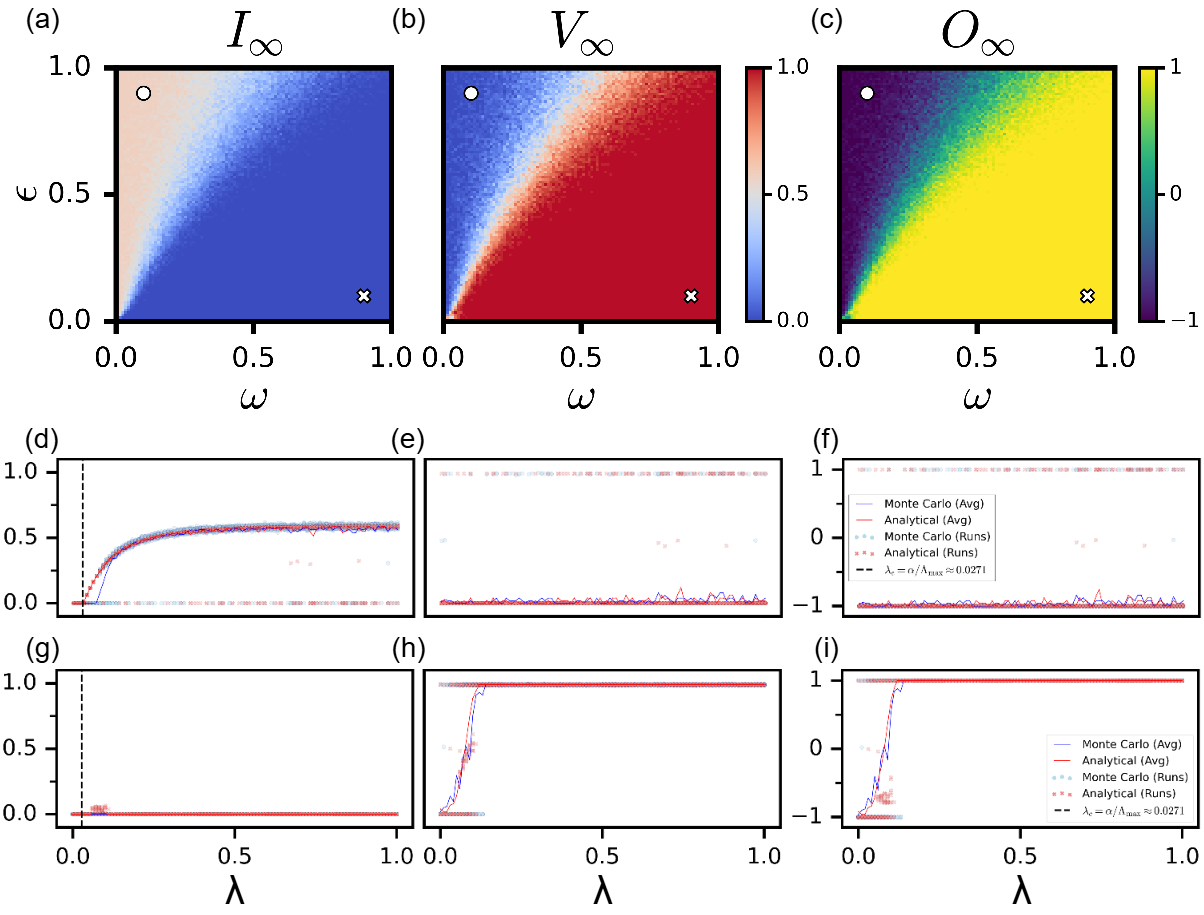}
	\caption{\textbf{Impact of peer influence ($\epsilon$) and risk perception ($\omega$) on steady-state outcomes.} 
		\textbf{Top row (a, b, c):} Heatmaps showing the steady-state fraction of infected individuals ($I_{\infty}$), vaccinated individuals ($V_{\infty}$), and average opinion ($O_{\infty}$) across the $(\omega, \epsilon)$ parameter space. Parameters are fixed at $\alpha=0.7$, $\phi=0.01$, and $\lambda=0.6$. Two distinct regimes are marked: the \textbf{white circle} represents a high-conformity, low-risk perception state $(\omega=0.2, \epsilon=0.8)$, while the \textbf{white cross} represents a low-conformity, high-risk perception state $(\omega=0.8, \epsilon=0.2)$.
		\textbf{Bottom rows:} Panels \textbf{(d, e, f)} and \textbf{(g, h, i)} show detailed sweeps for the parameter sets marked by the circle and cross, respectively. These panels display the steady-state metrics ($I_\infty$, $V_\infty$, $O_\infty$) from left to right as a function of the infection rate $\lambda$. 
		In the high-conformity regime (circle), strong peer influence suppresses the behavioral response, preventing effective vaccination and allowing infection to persist even at low $\lambda$. 
		In contrast, the high-risk perception regime (cross) shows that risk awareness drives a rapid shift to a pro-vaccine consensus ($O_\infty \approx 1$), effectively suppressing the epidemic across a wider range of transmission rates.}
	\label{fig:epsilon_omega_heatmaps}
\end{figure*}

To investigate how the steady-state average opinion ($O_{\infty}$), the final fraction of infected individuals ($I_{\infty}$), and the final fraction of vaccinated individuals ($V_{\infty}$) vary as the transmission rate $\lambda$ increases, we performed simulations using both MC simulation and a discrete-time formulation based on the microscopic Markov chain approach (MMCA)  (Fig.\  \ref{fig:steady_state_lines}). For each value of $\lambda \in [0,1]$ in increments of $0.01$, we carried out 50 independent runs to account for initial condition dependence. The MC outcomes are represented by blue dots with their mean shown as a blue solid line, while the MMCA results appear as red crosses with the corresponding mean plotted as a red solid line.
All simulations were conducted on a network consisting of $N = 2000$ nodes. The system initially contains $I_0 = 25$ infected nodes and is evolved for $10^4$ discrete time steps.


For Fig.\ \ref{fig:steady_state_lines}, we have considered the  recovery rate as $\alpha = 0.4$, and a waning-immunity rate for vaccinated individuals is fixed at $\phi = 0.01$. The opinion dynamics are driven by an initial fraction $D = 0.45$ of individuals with a positive bias toward vaccination, a risk-perception weight $\omega = 0.8$, and a peer-influence strength $\epsilon = 0.5$.
The initial opinions are set such that $D$ a fraction of individuals have opinions drawn uniformly from $[0,1]$ (pro-vaccine), and $1-D$ a fraction have opinions drawn uniformly from $[-1,0]$ (anti-vaccine). Infected individuals are chosen randomly at $t=0$. \\
The transitions observed in the infection profile is a dynamically meaningful feature rooted in the bifurcation structure of epidemic models. Classical compartmental models are well known to undergo various bifurcations at critical parameter thresholds, marking the boundary between disease extinction and sustained transmission. In the network setting, however, a full analytical bifurcation analysis becomes significantly more challenging due to heterogeneous degree distributions and the complex coupling between dynamical states across nodes. We therefore rely on mean-field approximations, which capture average system behavior but inherently smooth out stochastic fluctuations that are particularly prominent in finite-size networks. 
In our coupled opinion-epidemic framework, the critical transition point depends jointly on the dynamical parameters and the network structure, determining not only when infection takes hold but also when the population's collective response shifts from susceptible to infected. Precisely identifying this threshold provides a principled basis for designing targeted interventions, whether vaccination campaigns, behavioral nudges, or network-based strategies that keep the system below the critical point and suppress or prevent outbreak emergence entirely.


\subsection{Steady-States   vs.\  Infection Rate ($\lambda$)}
The average steady-state opinion of the whole community ($O_\infty$, Fig.\ \ref{fig:steady_state_lines} top panel):
For low $\lambda$ values, when the infection rate is minimal, opinions tend to be clustered around an anti-vaccine sentiment, reflecting minimal perceived risk. As $\lambda$ increases beyond the critical threshold, the average opinion shifts significantly towards a pro-vaccine stance. This indicates that a higher infection risk (driven by $\lambda$) strongly influences individuals' opinions, pushing them to favor vaccination due to higher risk perception (parameter $\omega$). Both MC and semi-analytical models show good agreement in this trend, indicating a robust socio-behavioral response to increasing epidemic threat. 
However, in the low infection rate regime, the analytical prediction (red line, averaged over 50 realizations) exhibits a near-sharp jump to a pro-vaccination state, when compared to the MC simulations (blue line; the discontinuity happens around 0.5). This discrepancy is likely due to finite-size effects, strong heterogeneity inherent in the network, and the presence of bistability in the stochastic dynamics. As seen in the time series analysis (see Appendix \ref{TS_dynamics}, Fig.\ \ref{fig:timeseries_infection}), the MC simulations in this regime capture a bifurcation where the system splits between disease-free and endemic states.
\par
Fraction of Infected Individuals ($I_{\infty}$, Fig.\ \ref{fig:steady_state_lines} middle panel):
At very low $\lambda$, the infection fraction $I_{\infty}$ is negligible, signifying that the epidemic cannot persist. As the $\lambda$ crosses a critical threshold ($\lambda_c \approx 0.0155$  for this parameter set), a gradual increase in  the $I_{\infty}$ is observed. 
The theoretical critical point $\lambda_c = \alpha / \Lambda_{\max}$ (Eq.~\eqref{eq:critical-lambda-sis}) is marked on the plot (dashed vertical line) and aligns well with the observed transition point in both simulation and analytical results. 
Beyond $\lambda_c$, $I_{\infty}$ continues to rise, but for higher $\lambda$, the opinion-driven vaccination plays a significant role in moderating the infection spread, preventing  
$I_{\infty}$
from reaching 1, demonstrating a form of self-limiting behavior driven by social response. \par
Fraction of Vaccinated Individuals ($V_{\infty}$, Fig.\  \ref{fig:steady_state_lines} bottom panel):
Corresponding to the increase in $I_{\infty}$ and the shift in opinion, the fraction of vaccinated individuals $V_{\infty}$ also shows a pronounced increase with $\lambda$. This is a direct consequence of higher risk perception (due to more infected neighbors) and peer influence, driving more susceptible individuals to vaccinate. The vaccination rate effectively acts as a control mechanism for the epidemic, dampening its potential severity. The trends are consistent across both modeling approaches, further solidifying the analytical model's reliability.

Furthermore, as shown in the supplementary material (Fig.\ \ref{fig:suppl_low_alpha}), the system interestingly exhibits a near-discontinuous transition in the low-$\alpha$ regime, where the population’s stance shifts abruptly from anti-vaccination to pro-vaccination.

\subsection*{$I_{\infty}$ as a function of  infection rates ($\lambda$) and recovery rates ($\alpha$). }

Figure \ref{fig:lambda_alpha_phase_plots} compares the phase diagrams of the steady-state infection density ($I_{\infty}$) obtained from MC simulations (left) and the semi-analytical model (right), across varying infection rates ($\lambda$) and recovery rates ($\alpha$). 
The dashed line denotes the theoretical critical threshold $\lambda_c = \alpha/\Lambda_{\max}$ (Eq.~\eqref{eq:critical-lambda-sis}). 
{As the recovery rate 
	$\alpha$ increases, the critical transmission threshold 
	$\lambda_c$ 
	also increases. The reddish and white regions together indicate the presence of infection, with the reddish area corresponding to higher infection levels. 
	The overall pattern of the infection regime, spanning both the reddish and white regions, remains qualitatively consistent between the simulation results and the theoretical predictions.
}


\subsection{Impact of peer influence ($\epsilon$)
	and risk perception ($\omega$)}
Figure~\ref{fig:epsilon_omega_heatmaps} illustrates the interaction between peer influence  
($\epsilon$)
and risk perception 
($\omega$) 
on the system’s steady-state behavior. The three panels respectively show the steady-state infection fraction (Fig.\ \ref{fig:epsilon_omega_heatmaps}(a)), vaccination fraction (Fig.\ \ref{fig:epsilon_omega_heatmaps}(b)), and average opinion (Fig.\ \ref{fig:epsilon_omega_heatmaps}(c)) for fixed parameters 
$\alpha = 0.7$ and $\lambda = 0.6$, over a sweep of $\epsilon$ and $\omega$ values.

In the left panel, we have reported the steady-state infection ($I_{\infty}$). 
High values of $\omega$ 
consistently lead to low infection levels, whereas regions with high peer influence and low risk perception exhibit significantly higher $I_{\infty}$.

When  $\omega$ is low, and  $\epsilon$ is high (white circle regime), the long-term infection level $I_{\infty}$ remains elevated. This occurs because ~$\sim55\%$ of the population begins with adverse opinions, which suppresses vaccination uptake (Fig.\  \ref{fig:epsilon_omega_heatmaps}(e)) and consequently sustains the infection (Fig.\ \ref{fig:epsilon_omega_heatmaps}(d)).
However, as the risk perception $\omega$ increases (moving toward the white cross in (Fig.\ \ref{fig:epsilon_omega_heatmaps}(a)), the population gradually shifts toward positive opinions (Fig.\ \ref{fig:epsilon_omega_heatmaps}(i)). This transition enhances the vaccination uptake (Fig.\ \ref{fig:epsilon_omega_heatmaps}(h)), leading to a reduction in the steady-state infection level (Fig.\ \ref{fig:epsilon_omega_heatmaps}(g)).
The vaccination fraction 
$V_{\infty}$ (middle panel) 
closely follows the trend of $O_{\infty}$ (right panel), reflecting the direct coupling between opinion and vaccination behavior (Eq.~\eqref{eq:vaccination-prob}). Regions with high $\omega$ and low
$\epsilon$ (white cross regime)
exhibit the strongest vaccination uptake, which in turn drives the reduction in $I_{\infty}$.
The average opinion tends toward strongly pro-vaccination values in regimes of high $\epsilon$ and $\omega$ (top right region). Strong peer influence accelerates consensus formation, while heightened risk perception biases this consensus toward vaccination when infection risks are salient (here, $\lambda = 0.6$).  


Overall, the results indicate that both $\epsilon$ and $\omega$ are key levers controlling collective epidemic outcomes. However, their influence operates differently: while $\omega$ (risk perception) drives the system toward a pro-vaccination state based on infection prevalence, $\epsilon$ (peer influence) regulates the system's sensitivity to that risk by reinforcing local consensus. For $\omega$ values moderately above 0.5, $\epsilon$ does not have a significant effect on the final outcomes, as the high risk perception becomes the dominant driver for vaccine uptake. In practice, risk perception alone can motivate protective behavior, but without sufficient peer coupling, such actions remain fragmented. Conversely, strong social conformity can amplify either cautious or complacent attitudes depending on prevailing norms. Hence, coordinated strategies that simultaneously enhance risk awareness and leverage social influence are most effective for steering the system toward widespread vaccination and epidemic suppression.

\subsection{Analytical Derivation of Saturated Vaccination Coverage}
\label{sec:analytical_saturation}

To understand the upper bound of vaccination coverage observed in our simulations (e.g., in regimes of high risk perception $\omega$ or low peer influence $\epsilon$), we derive an analytical expression for the steady state by following the flow of the microscopic dynamics.

We begin with the steady-state balance equation for the fraction of vaccinated individuals. For any node $j$, the rate of entering the vaccinated state must equal the rate of leaving it (waning immunity). The inflow is determined by the vaccination probability $\gamma_j^*$ acting on the susceptible fraction $S_j^* = (1 - V_j^* - I_j^*)$, while the outflow is determined by the waning rate $\phi$:
\begin{equation}
	\phi V_j^* = \gamma_j^* (1 - V_j^* - I_j^*).
\end{equation}
Rearranging to group the vaccination terms, we obtain:
\begin{equation}
	(\phi + \gamma_j^*) V_j^* = \gamma_j^* (1 - I_j^*).
\end{equation}
We substitute  the opinion-dependent vaccination probability, $\gamma_j^* = \frac{1 + O_j^*}{2}$, into the balance equation:
\begin{equation}
	\left(\phi + \frac{1 + O_j^*}{2}\right) V_j^* = \left(\frac{1 + O_j^*}{2}\right) (1 - I_j^*).
\end{equation}
Expanding both sides to separate the linear and non-linear terms:
\begin{align}
	\phi V_j^* + \frac{V_j^*}{2} + \frac{O_j^* V_j^*}{2} &= \frac{1}{2} - \frac{I_j^*}{2} + \frac{O_j^*}{2} - \frac{O_j^* I_j^*}{2}.
\end{align}
To obtain the macroscopic behavior, we sum this equation over all $N$ nodes in the network and normalize by $N$. Let $V_\infty = \frac{1}{N}\sum V_j^*$, $I_\infty = \frac{1}{N}\sum I_j^*$, and $O_\infty = \frac{1}{N}\sum O_j^*$. The summation yields:
\begin{equation}
	\phi V_\infty + \frac{V_\infty}{2} + \frac{\sum O_j^* V_j^*}{2N} = \frac{1}{2} - \frac{I_\infty}{2} + \frac{O_\infty}{2} - \frac{\sum O_j^* I_j^*}{2N},
	\label{eq:summed_balance}
\end{equation}

\begin{figure}
	\centering
	\includegraphics[width=0.5\linewidth]{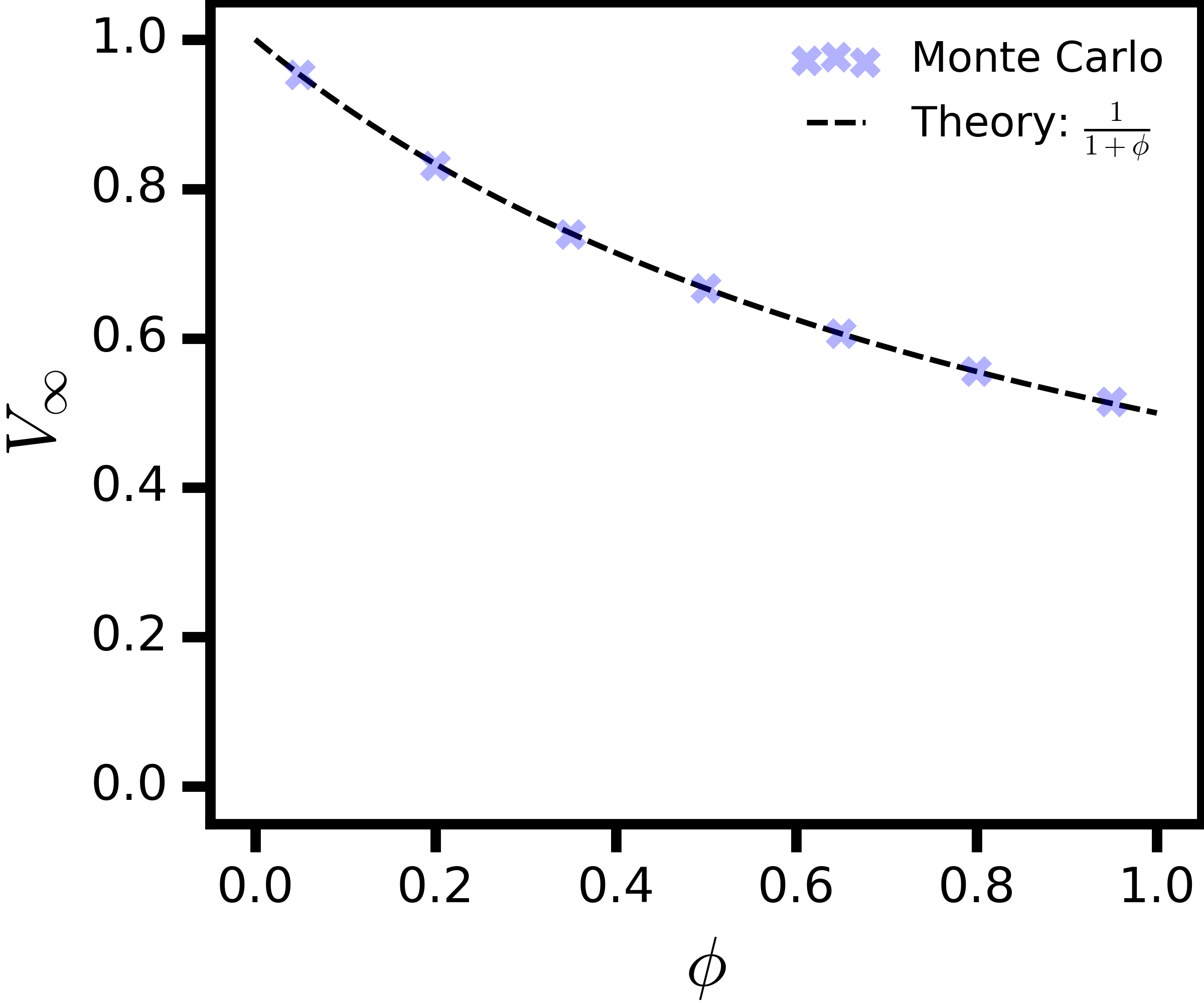}
	\caption{Verification of the analytical saturation coverage. The steady-state vaccination coverage $V_\infty$ is plotted against the waning rate $\phi$. The dashed line represents the theoretical prediction $V_\infty = \frac{1}{1+\phi}$ derived in Eq.~\eqref{eq:V_analytical_saturation}, while the points represent MC simulation results, showing excellent agreement. This was generated for fixed parameters $\alpha=0.1, \epsilon = 0.05, \omega = 0.9$ and $\lambda = 0.2$, with averages done over $100$ runs for each parameter set. 
	}
	\label{fig:phi_verification}
\end{figure}

For large risk perception $\omega$ (or setting small peer influence $\epsilon \to 0$), the presence of infection drives the opinion of all individuals toward the pro-vaccine maximum. As $O_j^* \to 1$, some terms  simplify as
$ \frac{\sum O_j^* V_j^*}{N} \approx V_\infty$, $\frac{\sum O_j^* I_j^*}{N} \approx I_\infty$, and $O_\infty \approx 1$. 
Substituting these limits into Eq.~\eqref{eq:summed_balance}:
\begin{align}
	\phi V_\infty + \frac{V_\infty}{2} + \frac{V_\infty}{2} &= \frac{1}{2} - \frac{I_\infty}{2} + \frac{1}{2} - \frac{I_\infty}{2} \nonumber \\
	\phi V_\infty + V_\infty &= 1 - I_\infty \nonumber \\
	V_\infty (1 + \phi) &= 1 - I_\infty.
\end{align}
Finally, we note that when $O_j^* \to 1$, the vaccination probability $\gamma_j^* \to 1$. Consequently, susceptible individuals vaccinate almost immediately, causing the effective infection probability to vanish ($I_\infty \to 0$). Setting $I_\infty = 0$ in the equation above yields the final saturation coverage:
\begin{equation} \label{eq:V_analytical_saturation}
	V_\infty = \frac{1}{1 + \phi}.
\end{equation}
This derivation confirms that in regimes where opinion consensus is reached ($O \approx 1$), the vaccination coverage is governed solely by the biological waning rate $\phi$, independent of the network topology or infection rate $\lambda$. In Fig.\ \ref{fig:phi_verification}, we see that MC simulations agree with the theoretical saturation coverage that we have derived in Eq. \ref{eq:V_analytical_saturation} and is consistent with the rectangular hyperbolic relationship between $V_\infty$ and $1+\phi$ expected. For a larger parameter sweep across various peer influence ($\epsilon$) and risk perception ($\omega$) values, refer to Fig.\ \ref{fig:suppl_phi_grid} in appendix \ref{Sat_Vac_Suppl}.

\section{Conclusion}
In this work, we have generalized a coupled opinion–epidemic model to operate on  network topologies, specifically Barab\'{a}si-Albert networks, which provide a more realistic representation of social structures than previous fully mixed population models. Our methodology combined both MC simulations for a stochastic agent-based approach and a semi-analytical discrete model (MMCA) for  deterministic understanding of the system dynamics.
We also derived analytical expressions for critical infection thresholds, which were numerically validated through our simulations, showing strong agreement with the theoretical predictions. Our extensive simulations and analytical results confirm phase transitions 
in the coupled system, phenomena previously observed in fully connected networks \cite{pires2018sudden} but now extended and validated in heterogeneous environments. 
A key aspect of our investigation was to quantify the impact of peer influence ($\epsilon$) and risk perception ($w$) on the steady-state outcomes of infection prevalence, average public opinion, and vaccination rates. Our findings clearly demonstrate how variations in these parameters, alongside the infection transmission rate ($\lambda$), 
significantly alter the system's behavior, leading to different epidemic sizes and opinion distributions. The consistent agreement between the MC simulations and the semi-analytical model throughout these investigations shows the robustness and reliability of our discrete modeling framework.
{This coupled opinion-infection model in complex networks will   enable us to understand  how opinions shift in response to risk perception and peer influence The analysis can suggest  strategies to counter vaccine hesitancy and misinformation by leveraging network dynamics.}  Identifying critical thresholds and parameter sensitivities can aid in designing more effective and targeted vaccination campaigns, focusing resources on specific network components or social influence mechanisms\cite{wang2023subsidy}.
The framework also offers a robust tool for exploring socio-epidemic dynamics in environments that accurately reflect the complex connectivity of human populations.
This work provides a basis for data-driven interventions by deepening our understanding of how social factors interact with biological processes to shape public health outcomes through theoretical models and realistic network structures.

 While the Barabási-Albert networks employed in this study effectively approximate the structural heterogeneity of social systems, future research could utilize real-world topologies to further validate these dynamics. To capture a more nuanced behavioral evolution, the model could incorporate mechanisms for opinion decay over time or in response to prolonged periods of low infection prevalence. Furthermore, the current decision-making logic could be extended by integrating evolutionary game theory, allowing individuals to  choose to vaccinate based on a strategic comparison of the costs associated with vaccination versus the perceived risks of infection. Another promising direction involves introducing agent-level heterogeneity to explore how individual-specific behavioral traits and social differences diversify vaccination uptake and epidemic spread. Finally, future work could also investigate the impact of varying waning rates ($\phi$) to reflect the specific clinical characteristics and biological properties of different diseases. The analytical results presented in this work are derived within specific limiting regimes of transmission rate, a complete and unified analytical treatment of all observables encompassing the full parameter space and arbitrary network topologies remains an important and natural extension of the current framework.



\section{Appendix}
\subsection{Time Series Dynamics}
\label{TS_dynamics}

\begin{figure*}[ht]
	\centering
	\includegraphics[width=0.8\textwidth]{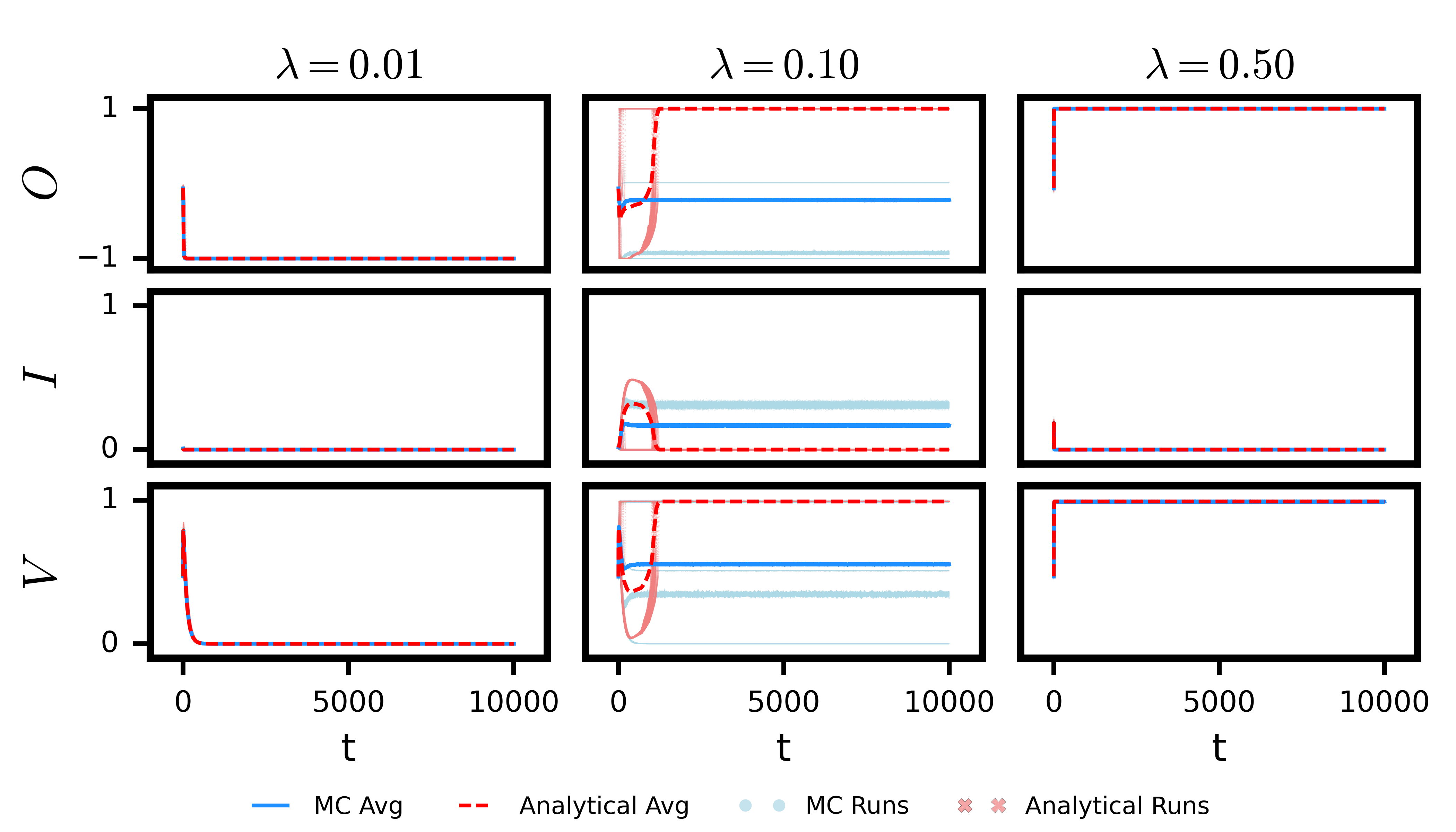}
	\caption{\textbf{Time evolution of the fraction} of infected individuals ($I(t)$), vaccinated individuals ($V(t)$) and opinions ($O(t)$) for different values of the infection rate $\lambda$, for $\omega=0.8$, $\epsilon=0.5$, $\alpha=0.4$, $\phi=0.01$. 
		Individual MC runs are shown as blue dots, MC averages in solid blue, and semi-analytical runs are shown as red crosses with the average as dashed red lines. The averages have been done across runs at each timestep (50 individual realizations for each of MC and analytical model). The plots demonstrate the convergence of both models to steady-state values, with the presence of bistability in the MC simulations for specific values of $\lambda$.}
	\label{fig:timeseries_infection}
\end{figure*}

Figure~\ref{fig:timeseries_infection} presents the time evolution of the system's key metrics: average opinion ($\langle O \rangle$), fraction of infected individuals ($\langle I \rangle$), and fraction of vaccinated individuals ($\langle V \rangle$); for three distinct infection rates $\lambda = 0.01, 0.10, \text{ and } 0.50$. These trajectories illustrate the coupled system's convergence to steady state and highlight critical differences between the stochastic and semi-analytical approaches. The values of $\lambda$ have been specifically chosen from various phase transition zones of Fig.\ \ref{fig:steady_state_lines} to visualize the dynamics of each zone more closely. For low infection rates ($\lambda=0.01$), both models rapidly converge to a disease-free equilibrium with low vaccination coverage, driven by the absence of sustained infection risk. Conversely, at high infection rates ($\lambda=0.50$), the high perceived risk drives a rapid consensus toward pro-vaccine opinions ($\langle O \rangle \approx 1$), leading to high vaccination coverage that suppresses the infection to a low endemic level. In both these limiting regimes, the deterministic semi-analytical model (dashed red lines) shows agreement with the average of the Monte Carlo (MC) simulations (solid blue lines). However, in the intermediate regime ($\lambda=0.10$), a notable discrepancy emerges between the two approaches. The MC simulations exhibit bistability with individual realizations (light blue points) splitting between a disease-free state ($\langle I_\infty\rangle \sim 0$ and an endemic state ($\langle I_\infty\rangle \sim 0.33$). The semi-analytical model however converges to a single, disease-free fixed point. This divergence highlights a key limitation of our semi-analytical framework. 
As a result, the global MC average is ``pulled up" by some endemic events, appearing higher than the deterministic prediction which effectively tracks only the disease-free branch of the bistable system. This stochastic bifurcation provides a direct explanation for the quantitative mismatch observed between the semi-analytical curves and MC averages in the intermediate infection regime. We have observed such a gap in Fig.\ \ref{fig:steady_state_lines} previously. 




\subsection{Saturation of Vaccinated individuals}
\label{Sat_Vac_Suppl}

\begin{figure*}[ht]
	\centering
	\includegraphics[width=0.95\textwidth]{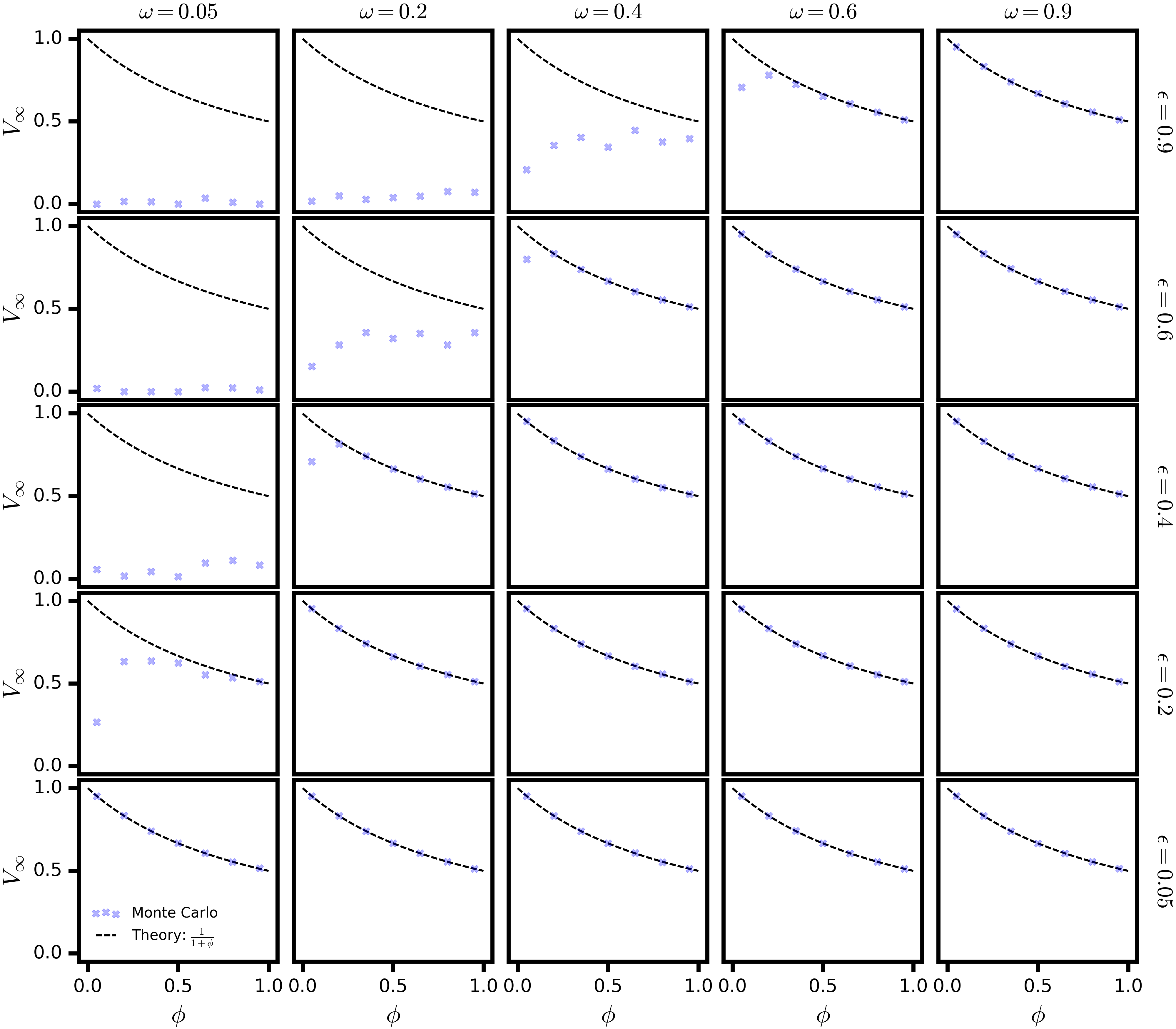} 
	\caption{\textbf{Extended verification of analytical vaccination saturation across parameter space.} Steady-state vaccination coverage $V_\infty$ as a function of waning rate $\phi$ for a parameter sweep of risk perception ($\omega \in \{0.05, 0.4, 0.9\}$) and peer influence ($\epsilon \in \{0.05, 0.4, 0.9\}$). The analytical prediction $V_\infty = 1/(1+\phi)$ (dashed line) remains robust across the parameter space, particularly in regimes where opinion reaches a pro-vaccine consensus ($\omega >> \epsilon$ and/or low $\epsilon$). In regimes of very low risk perception and high peer influence (top-left), the simulation results deviate as the system fails to reach the saturation regime assumed in Eq.~\eqref{eq:V_analytical_saturation}.}
	\label{fig:suppl_phi_grid}
\end{figure*}

This section provides an extended verification of the analytical saturation coverage derived in Eq.\ \eqref{eq:V_analytical_saturation}. As illustrated in Fig.\ \ref{fig:suppl_phi_grid}, the theoretical prediction $V_{\infty} = 1/(1+\phi)$ remains remarkably robust across a broad parameter sweep of risk perception ($\omega$) and peer influence ($\epsilon$). This saturated state occurs specifically in regimes where the population reaches a pro-vaccine consensus ($O_{\infty} \approx 1$), effectively making vaccination coverage independent of network topology or infection rates, and instead governed solely by the waning rate $\phi$ under certain peer influence and risk perception values.

However, the Monte Carlo simulation results deviate from the theoretical curve in the top-left sub-panels of Fig.\ \ref{fig:suppl_phi_grid} (e.g., $\omega=0.05$ and $\epsilon=0.9$). In these specific regimes, the combination of very low risk awareness and high social conformity prevents the system from escaping an anti-vaccine state. The theoretical prediction remains robust in a substantial part of the parameter space.

\subsection{Transitions in the low-recovery regime}
\begin{figure*}[ht]
	\centering
	\includegraphics[width=0.6\textwidth]{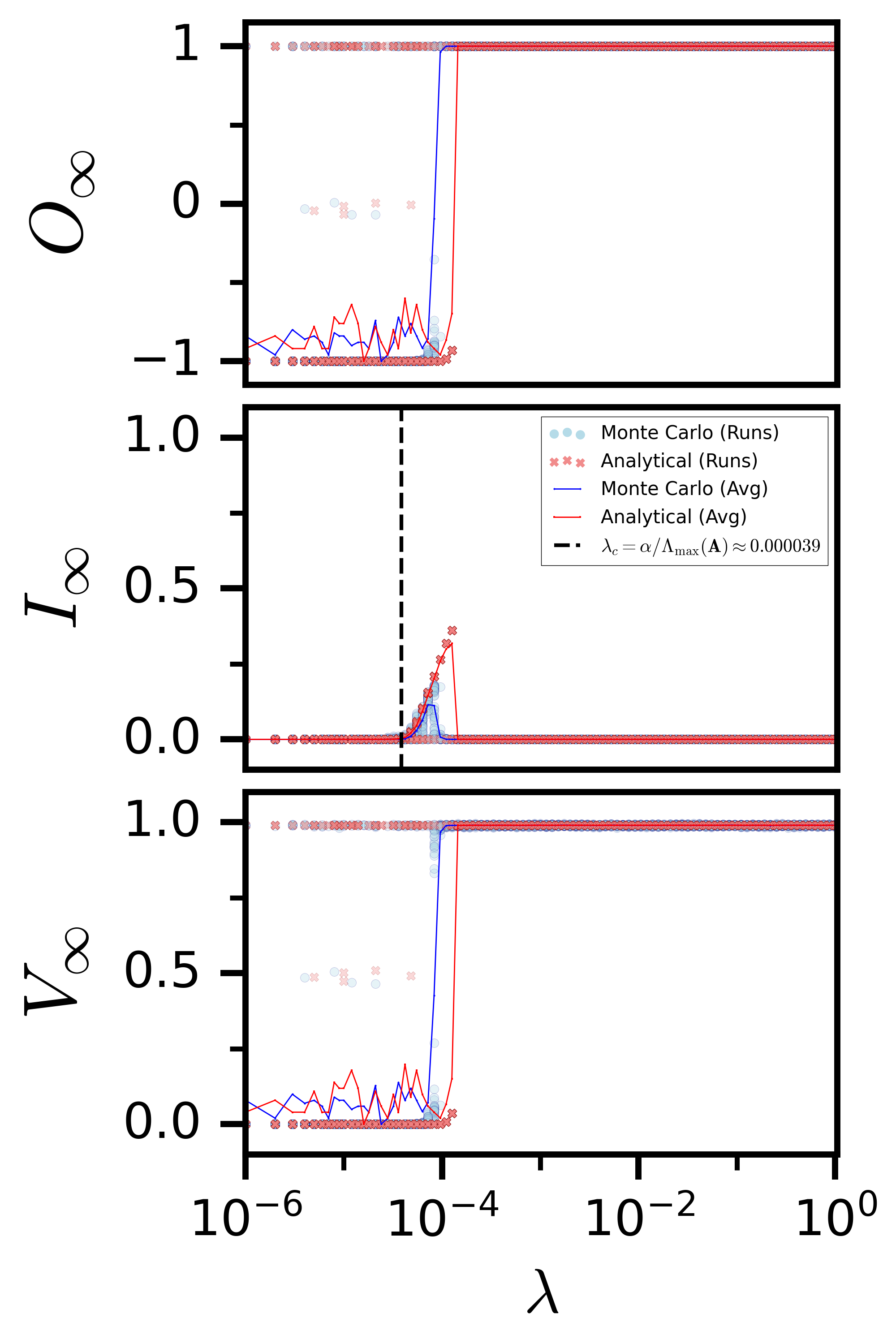}
	\caption{\textbf{Steady-state metrics for low recovery rate ($\alpha = 0.001$)}. Remaining fixed parameters are $\phi=0.01, \epsilon=0.5, \omega=0.8$. Final values for opinion ($O_\infty$), infection ($I_\infty$), and vaccination ($V_\infty$) are shown across varying infection rates $\lambda$. The dashed line in the middle panel indicates the theoretical critical threshold $\lambda_c \approx 0.000039$. In the lower recovery rate regime, we see a near-discontinuous transition of opinions from an anti-vaccine to pro-vaccine stance.}
	\label{fig:suppl_low_alpha}
\end{figure*}
To highlight the dynamics across several orders of magnitude, the transmission rate $\lambda$ is plotted on a log-scale. This scaling reveals a sharp-like transition in public sentiment ($O_{\infty}$). Because the low recovery rate allows the infection to persist more easily even at low transmission rates, the perceived risk accumulates rapidly once $\lambda$ crosses the threshold. This results in a near-discontinuous flip in opinion from a nearly universal anti-vaccine stance ($O_{\infty} \approx -1$) to a strong pro-vaccine consensus ($O_{\infty} \approx 1$). This behavioral transition consequently drives a rapid increase in vaccine uptake ($V_{\infty}$) to suppress the epidemic in this high-sensitivity regime.  

\subsection{Role of hubs (leaders) in opinion model}

\begin{figure*}
    \centering

    \begin{minipage}{0.48\textwidth}
        \centering
        \includegraphics[width=\linewidth]{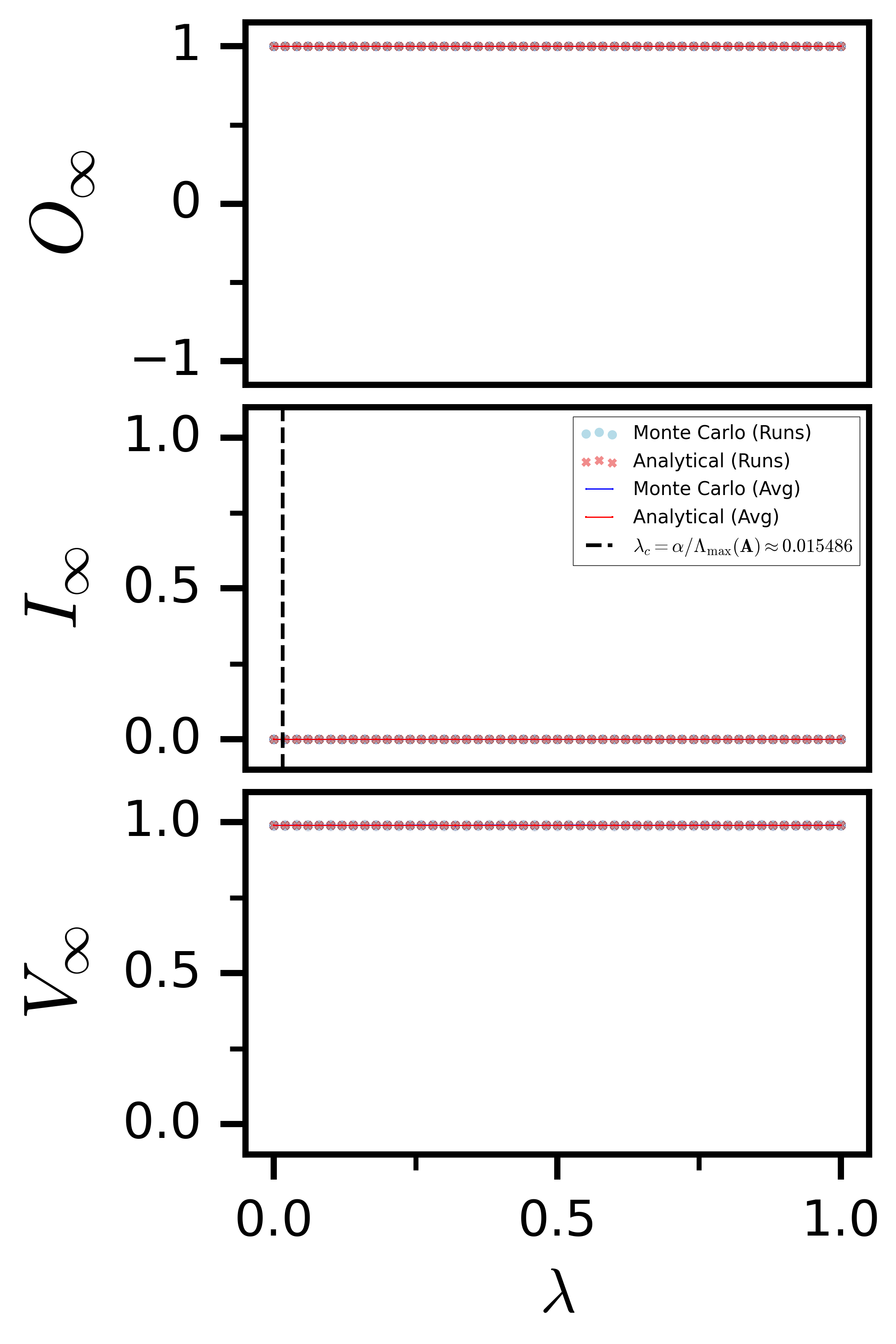}
    \end{minipage}
    \hfill 
    \begin{minipage}{0.48\textwidth}
        \centering
        \includegraphics[width=\linewidth]{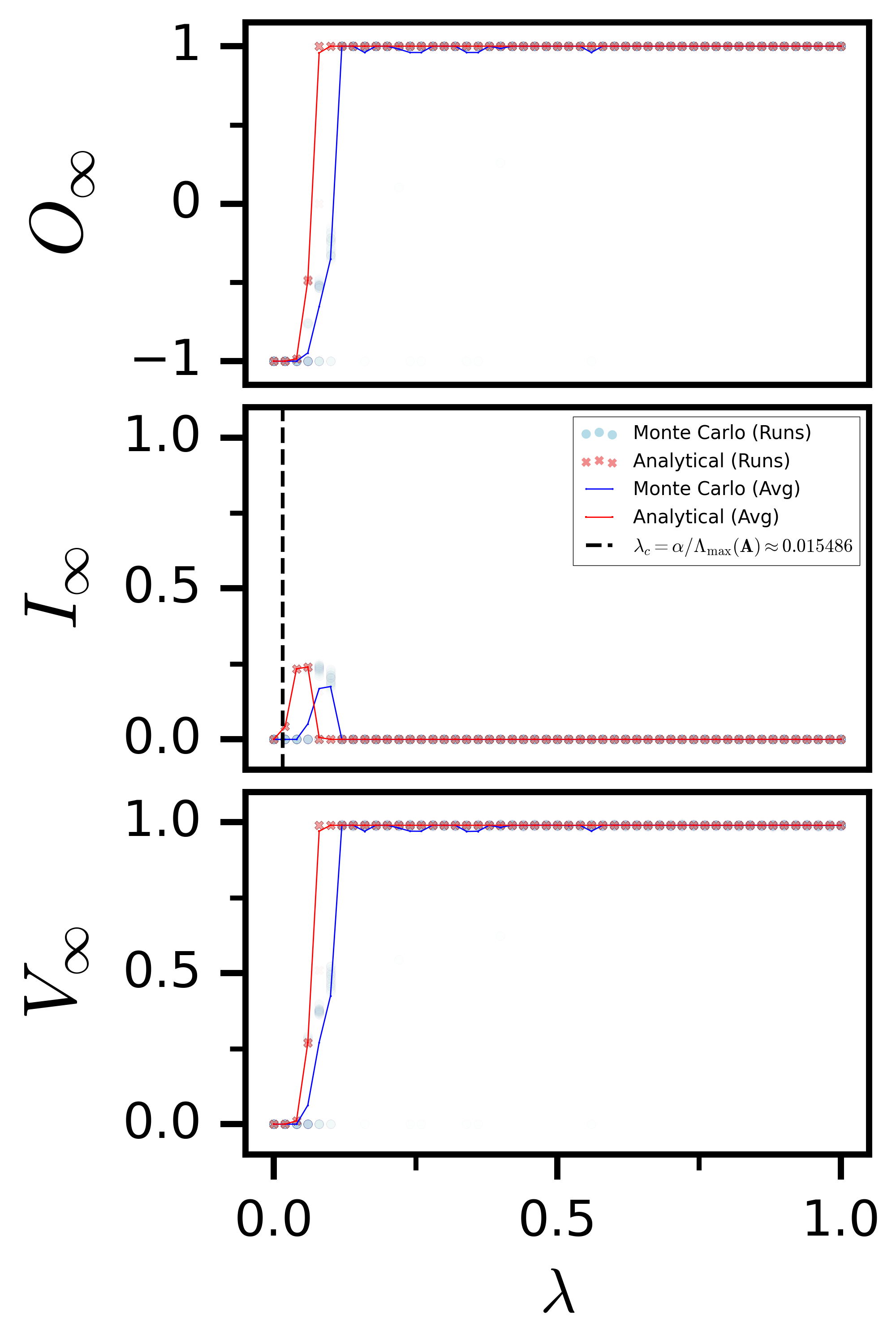}
    \end{minipage}

    \vspace{10pt}
    
    \caption{\textbf{Steady-state metrics under hub-weighted influence} on a BA network ($N=2000$, $\langle k \rangle \approx 4$) with biased leaders ($\alpha=0.4, \epsilon=0.8, \omega=0.2, \phi=0.01$). 
    Left panel shows the system behavior when the top 10 highest-degree nodes are forced to a positive initial opinion, leading to a consensus of $O_{\infty} \approx 1$ and high vaccination rates across all $\lambda$. 
    Right panel illustrates the system when hubs are initialized with negative opinions, showing a clear phase transition where the system only reaches positive consensus after the infection rate $\lambda$ becomes high enough to overcome the initial negative bias of the leaders. 
    The semi-analytical model (red lines) maintains high accuracy in predicting these leader-driven dynamics.}
    \label{fig:hub_leader_effects}
\end{figure*}

We further investigated the effect of hub nodes considering them as leaders by relaxing the assumption of equal weights in the opinion update process. In this weighted model, we define the ``influence capacity'' of a node as $\psi_{j} = k_j / k_{max}$. The influence of neighbor $j$ on node $i$ is scaled by this factor, reflecting real-world scenarios where the influence of hubs dominates over the influence of lower-degree or peripheral nodes. Simulations on the standard BA graph (Fig.~\ref{fig:hub_leader_effects}) show that the nature of steady-state transitions for opinion ($O_{\infty}$), infection ($I_{\infty}$), and vaccination ($V_{\infty}$) remain consistent with the unweighted case. The modified opinion update equation, incorporating the influence capacity, is given by:

\begin{equation}
	O_i(t+1) = O_i(t) 
	+ \frac{\epsilon}{K_i}\sum_{j=1}^N A_{ij} \psi_j O_j(t) 
	+ \frac{\omega}{K_i}\sum_{j=1}^N A_{ij} I_j(t).
	\label{eq:opinion-update-weighted}
\end{equation}

The simulation results indicate that the initial state of hub nodes, functioning as ``leaders," impacts the system's global steady state. As shown in Fig.~\ref{fig:hub_leader_effects}, initializing the top $0.5\%$ highest-degree nodes with positive opinions ($O_i \in [0, 1]$) drives the network toward a pro-vaccination consensus ($O_{\infty} \approx 1$) across the full range of $\lambda$. Under this configuration, the weighted influence capacity $\psi_j$ ensures near-universal vaccination ($V_{\infty} \approx 1$), maintaining the epidemic fraction ($I_{\infty}$) near zero even at low infection rates.

In contrast, initializing these hubs with negative opinions ($O_i \in [-1, 0]$) inhibits vaccination uptake. In the low-$\lambda$ regime, the negative bias of the leaders dominates the network, resulting in $O_{\infty} \approx -1$ and $V_{\infty} \approx 0$. A phase transition occurs only at a relatively high infection rate, where the infection-driven opinion term ($\omega$) surpasses the negative influence of the leaders, shifting the steady state toward positive consensus.

The semi-analytical model remains robust under heterogeneous weighting and biased initialization. The close alignment between Monte Carlo and semi-analytical results confirms the validity of our approach.

\subsection{Results on real-world online social networks}
\label{real-world}
\begin{figure*}
    \centering

    \begin{minipage}{0.48\textwidth}
        \centering
        \includegraphics[width=\linewidth]{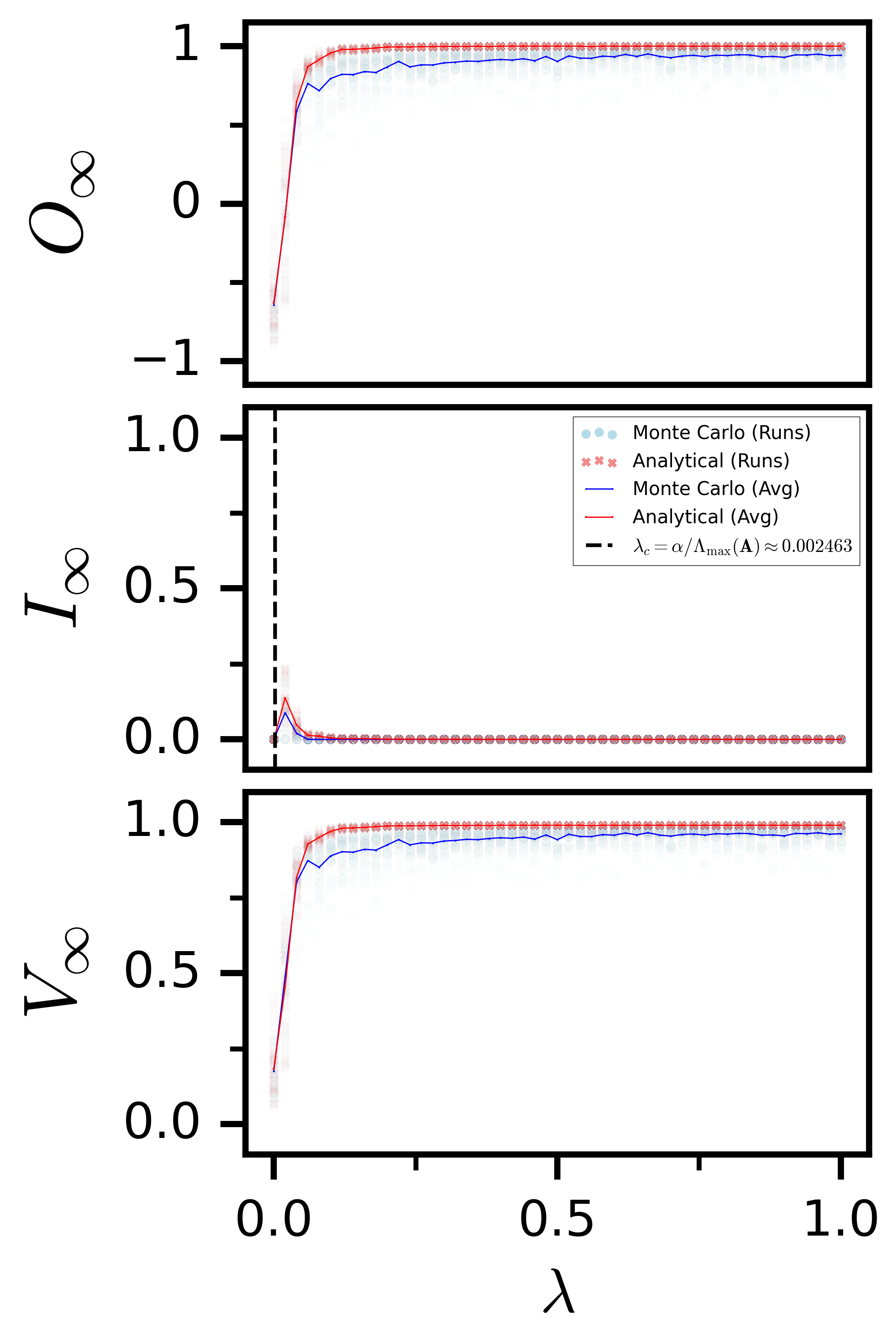}
    \end{minipage}
    \hfill 
    \begin{minipage}{0.48\textwidth}
        \centering
        \includegraphics[width=\linewidth]{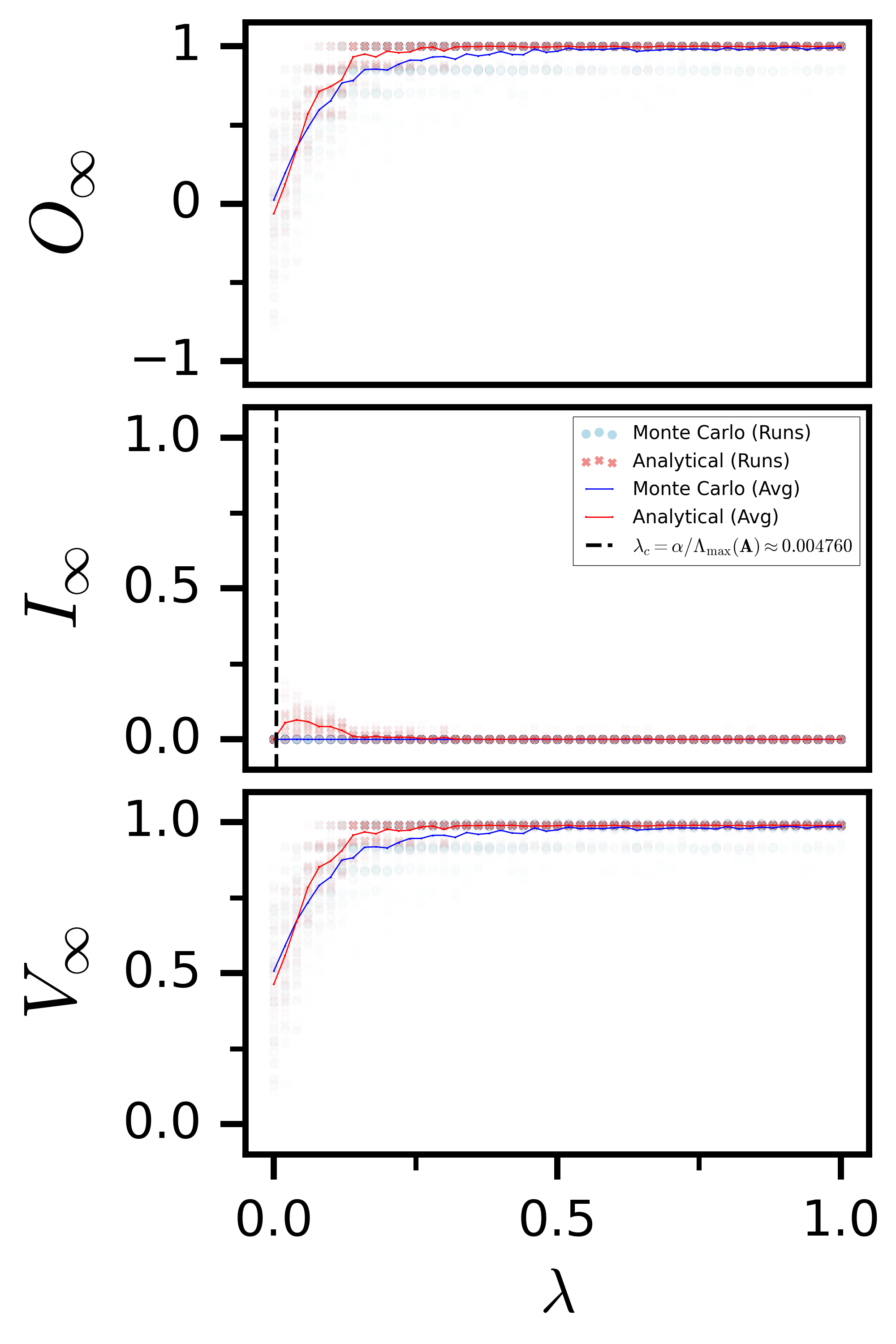}
    \end{minipage}

    \vspace{10pt} 
    
    \caption{Steady-state average opinion ($O_{\infty}$), fraction infected ($I_{\infty}$), and fraction vaccinated ($V_{\infty}$) as functions of the infection rate $\lambda$ ($\alpha = 0.4, \epsilon = 0.5, \omega = 0.8, \phi = 0.01$). Left: Results for the Facebook graph with node $N=4039$ and $\langle k \rangle=43.6910$. Right: Results for the Epinion graph with node $N=467$ and $\langle k \rangle=28.5482$. Plotted figures demonstrate that the phase transition features observed in standard network models remain robust when extended to real-world network structures. Furthermore, the semi-analytical model demonstrates strong agreement with the Monte Carlo simulations, with the critical transition value approaching zero, a direct consequence of the inherent heterogeneity present in these networks.}
    \label{fig:steady_state_real_networks}
\end{figure*}

To further validate the robustness and applicability of our results, we extended our approach to two well-known online social networks: a Facebook network with $N=4039$ nodes and mean degree $\langle k \rangle=43.6910$, and an Epinions network with $N=467$ nodes and mean degree $\langle k \rangle=28.5482$, demonstrating that our framework generalizes effectively beyond synthetic network settings. As illustrated in Fig.~\ref{fig:steady_state_real_networks}, the coupled dynamics of opinion and epidemic spreading remain qualitatively similar on these real-world topologies. Despite the presence of complex local bottlenecks and community structures, the system still undergoes a transition to a pro-vaccine consensus ($O_{\infty} \approx 1$) as $\lambda$ increases. We observe close alignment between the stochastic Monte Carlo runs and the semi-analytical MMCA predictions on these datasets. Real-world networks are inherently highly heterogeneous, and as a consequence of their large spectral radius, the critical transition value tends toward zero, as clearly illustrated in Fig.~\ref{fig:steady_state_real_networks}.

\color{black}

\bibliography{sample}

\section*{Data availability}
No new data were generated or analyzed in this study. All results were obtained from the proposed mathematical model through computer simulations. The code used to support the findings of this study is available from the corresponding author upon reasonable request.

\section*{Funding}
This research received no external funding.
\subsection*{Acknowledgements}
The authors thanks Michael Small for useful discussions. CH acknowledges support from ARNF India (Grant Number ANRF/ECRG/2024/000207/PMS). TK and SG acknowledge support from the National Science Centre, Poland, OPUS Programs (Project No. 2021/43/B/ST8/00641).

\section*{Author contributions statement}
A.R. and U.S. performed the Monte Carlo and numerical simulations, conducted the formal analysis, investigated the results, and wrote the main manuscript text. S.G. and C.H. contributed to the conceptualization and methodology of the study, supervised the investigation, validated the results, and participated in writing the original draft as well as reviewing and editing the manuscript. T.K. supervised the work and contributed to the review and editing of the manuscript.
\section*{Competing interests}
The authors declare no competing interests.

\section*{Additional information}

\end{document}